\documentclass[journal]{IEEEtran}

\usepackage{citesort}

\usepackage{amsmath}
\usepackage{epsfig}
\usepackage{subfigure}
\usepackage{enumerate}

\usepackage{booktabs}
\usepackage{multirow}
\usepackage{bigstrut}

\usepackage{amssymb}

\begin{document}
\title{An Empirical Performance Study of Intra-vehicular Wireless Sensor Networks under WiFi and Bluetooth Interference}

\author{\authorblockN{Jiun-Ren Lin$^1$, Timothy Talty$^2$, and Ozan K. Tonguz$^1$}\\
\authorblockA{$^1$Carnegie Mellon University, ECE Department, Pittsburgh, PA 15213-3890, USA\\
$^2$General Motors LLC, ECI Lab, Research and Development, Warren, MI 48092-2031, USA\\
Email: j.lin.us@ieee.org, timothy.talty@gm.com, tonguz@ece.cmu.edu}
}%
\maketitle

\begin{abstract}
Intra-Vehicular Wireless Sensor Network (IVWSN) is a new automotive architecture that applies wireless technologies to the communications between Electrical Control Units (ECUs) and sensors. It can potentially help achieve better fuel economy, reduce wiring complexity, and support additional new applications. In the existing works, most of the popular wireless technologies applied on IVWSNs occupy the same 2.4 GHz ISM frequency bands as WiFi and Bluetooth do. It is therefore essential to evaluate the performance of IVWSNs under interference from WiFi and Bluetooth devices, especially when these devices are inside the vehicle.
In this paper, we report the results of a comprehensive experimental study of IVWSNs based on ZigBee and Bluetooth Low Energy under WiFi and Bluetooth interference. The impact of the interference from Bluetooth and WiFi devices can be clearly observed from the experiments. The results of the experiments conducted suggest that Bluetooth Low Energy technology outperforms ZigBee technology in the context of IVWSNs when WiFi interference exists in the car. 
\end{abstract}


\IEEEpeerreviewmaketitle

{\small \textsl{\textcopyright2013 IEEE. Personal use of this material is permitted. Permission from IEEE must be
obtained for all other uses, in any current or future media, including
reprinting/republishing this material for advertising or promotional purposes, creating new
collective works, for resale or redistribution to servers or lists, or reuse of any copyrighted
component of this work in other works.}}

\section{Introduction}
\begin{figure}[tbp]
\centering
\includegraphics[scale = 0.56, trim = 5mm 5mm 5mm 5mm, clip]{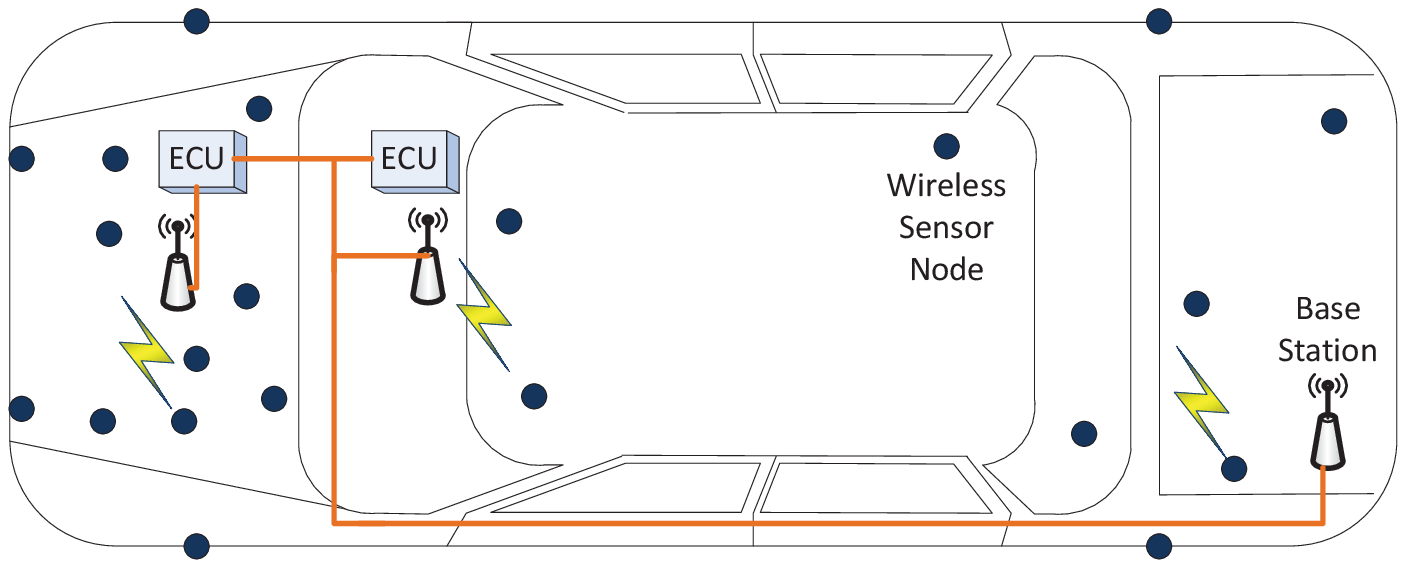}
\caption{An intra-vehicular wireless sensor network}
\label{SampleWSN}
\end{figure}

Modern vehicles are highly computerized, and most of the vehicular operations are controlled by sophisticated embedded systems. As more features are added to the vehicles, the number of vehicular sensors keeps increasing. Currently, almost all of the sensors inside a vehicle connect to their destination Electronic Control Units (ECUs) through wired connections. The increasing number of sensors leads to more wires that have to be added into the vehicles. These additional wires raise the complexity of the vehicles and increase the cost and complexity for car manufacturers to design and assemble the vehicles. They also contribute to the weight of the vehicles, thus limiting the range of possible positions for installing sensors. 
Due to these reasons, Intra-Vehicular Wireless Sensor Networks (IVWSNs) have recently received a great deal of attention in the automotive industry~\cite{GM-intra-vehicular}. By utilizing wireless technologies, IVWSNs can possibly help vehicle manufacturers reduce the design and assembly cost, achieve better fuel economy and performance, and support new applications~\cite{JRVTC2011}. An example architecture of IVWSNs is shown in Figure~\ref{SampleWSN}.

Since most of the wireless sensors in IVWSNs will be powered by batteries and do not require a high data rate, low-power wireless technologies, such as ZigBee/IEEE 802.15.4 or Bluetooth Low Energy (BLE), are good candidates to be used as the physical (PHY) and MAC layer protocols~\cite{ZigBee-based-WSN}\cite{2012JR_BLE_Tech}. Both of the two wireless technologies operate in 2.4 GHz ISM band. In the mean time, some car manufacturers currently have built-in WiFi as an option on their production cars (e.g., autonet~\cite{autonet}), and most of them provide integrated Bluetooth handsfree systems. WiFi (i.e., IEEE 802.11b/g) and Bluetooth also operate in the same 2.4 GHz ISM band, and therefore the coexistence problem is significant for IVWSNs. Another emerging source of WiFi signal is from personal hotspots, which enable users to share their cellular Internet access to other WiFi devices. This implies that unexpected WiFi interference can be also introduced from mobile devices.

The main objective of this paper is to study the impact of WiFi and Bluetooth interference on the performance of IVWSNs. While there are a number of existing papers that study the impact of such interference, most studies focus on investigating the effect of interference in an office or indoor environment. For instance, Shin et al. conducted a packet error rate analysis of ZigBee under WiFi and Bluetooth interference~\cite{Shin2007}. Shuaib et al. performed some experiments to study the performance degradation of a ZigBee wireless sensor network, Bluetooth, and WiFi devices, when they co-exist in an office environment~\cite{Shuaib2006}. Sikora and Groza reported experimental results on ZigBee performance under interference from WiFi (802.11b only), Bluetooth, and a microwave oven~\cite{Sikora05}. Chong et al. reported analytical results on the throughput of ZigBee network when WLAN interference exists~\cite{chong2007an}. Besides the aforementioned works, an interesting work done by Francisco et al. empirically investigates the impact of Bluetooth interference on ZigBee wireless sensor networks in a static setting where the experiment was performed when a car is parked in a parking lot~\cite{Francisco2009}. In this paper, we conduct a comprehensive experimental study that investigates the impact of both WiFi and Bluetooth interference in two different types of IVWSNs, and the experiments are conducted in both parking and moving scenarios.

The rest of this paper is organized as follows. Section II provides the details of the experimental platform. Section III introduces the experimental setup. Section IV depicts the experimental results and the major observations made in the experiments. 
Finally, concluding remarks are given in Section V.

\section{Experimental Platform}

In this study, two different wireless technologies (i.e., ZigBee and BLE) are used as the underlying platforms for IVWSNs and each platform is evaluated under two different kinds of interference (i.e., WiFi and Bluetooth). Details of each of the four components are described in the following subsections.

\subsection{IVWSN: Bluetooth Low Energy}
The Texas Instruments CC2540 mini development kit~\cite{CC2540kit} is used to develop the experimental platform for the BLE-based IVWSNs. This platform consists of three BLE devices: one master device, one slave device, and one packet sniffer. 
After booting up, the slave device will broadcast advertisements. Once the master device hears the advertisements, it will establish a connection with the slave. After the connection is established, there will be periodic packet exchanges (called \textit{connection events}) between the master and the slave device. The interval of the connection events is set as 0.25 second. 

\begin{figure}[tbp]
\centering
\includegraphics[scale = 0.45]{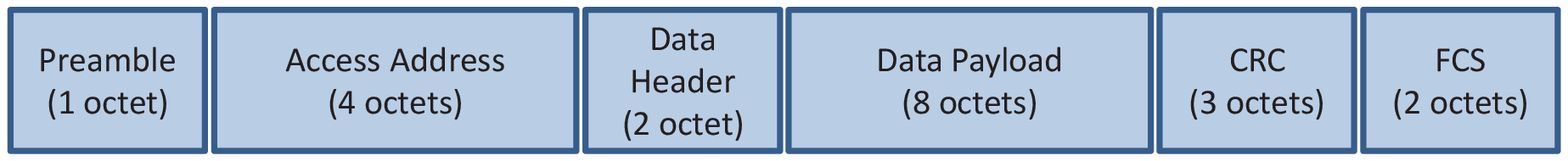}
\caption{The packet format of the BLE platform}
\label{BLEFrameFormat}
\end{figure}

In the experiments, in order to evaluate the system performance under interference, it is desired to create a constant packet flow from the slave to the master device, and all of the slave-to-master packets have to be of the same length. However, due to the limitations of the development kit, i.e., the BLE protocol stack is provided as object codes, we have no direct control to the PHY and MAC layers of the BLE stack and hence cannot modify the packet/frame format directly. Our solution is to enable the notification feature of the Generic Attribute Profile (GATT)~\cite{BLEspec}. After the notification of a specific characteristic (i.e., a data field in a GATT profile) on the slave is enabled by the master, every time when the characteristic is changed, the slave will generate a notification packet to the master. The notification is not necessary to be included in a connection event, but if the time of the characteristic change is very close to next connection event, the notification will be combined into the next connection event packet. Therefore, the application layer on the slave device is programmed to modify the characteristic value every 0.25 seconds, and then we can observe that every connection event packet from the slave to the master includes the notification, and the data payload is fixed at 8 bytes, which is a typical data size for IVWSNs. The packet format of the connection event packets from the slave to master is shown in Figure~\ref{BLEFrameFormat}. The total length of a packet including a 1-byte preamble is 20 bytes. The transmit power is 0 dBm.

Moreover, because no direct control to the PHY layer is provided by the development kit, a BLE packet sniffer is used to overhear the packets in order to measure the RSSI and calculate the Packet Error Rate (PER) and Packet Reception Rate (PRR). 
In each experiment, the packet sniffer is used to capture the BLE packets, and it stores the received packets to a binary log file, which is offline parsed to collect the packet information and calculate the statistic. Note that we only collect the information of the connection event packets sent from the slave to master\footnote{This is because the uplink traffic is sensor data and is the traffic of interest in an IVWSN.}, and other packet types can be distinguished and discarded according to the packet header.

In addition, an u-blox EVK-6P GPS evaluation kit is used to collect the GPS information at a rate of 4 Hz when the sniffer captures the BLE packets. The GPS information is synchronized with the captured BLE packets to provide timestamps and other information (e.g., vehicle speed and location).

\subsection{IVWSN: ZigBee}
The ZigBee IVWSN platform consists of two FireFly sensor nodes~\cite{FireFly} --- one transmitting node and one receiving node. The FireFly sensor node has a Texas Instruments CC2240 RF chip and it is compliant with the PHY and MAC protocols of Zigbee (and thus IEEE 802.15.4 standard). However, in this paper, we only use standard PHY of IEEE 802.15.4 on the Zigbee platform. Since there is only one pair of Zigbee devices in the experiments, we apply a simple TDMA MAC protocol.  

\begin{figure}[tbp]
\centering
\includegraphics[scale = 0.34]{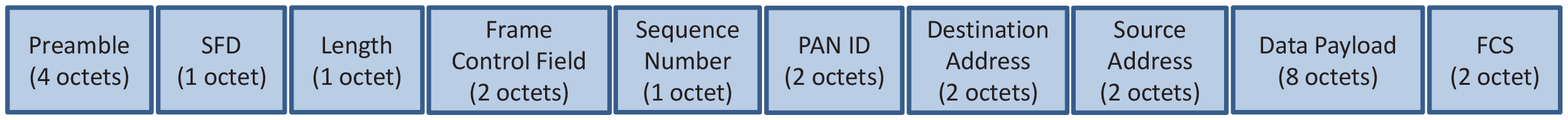}
\caption{The packet format of the ZigBee platform}
\label{ZigBeeFrameFormat}
\end{figure}

Similar to the BLE-based IVWSNs, the same 8-byte data payload is used to represent the sensor data. However, due to a larger packet header (see Figure~\ref{ZigBeeFrameFormat}), the size of a ZigBee packet is 25 bytes (including a 4-byte preamble). The Zigbee packets are transmitted at a rate of 4 Hz using channel 17 (2435 MHz) and transmitting power of 0 dBm.

The Zigbee packet receiver is connected to a PC through a serial link. There are two programs on the PC; one is the program which records the received packet information from the receiver and the corresponding GPS information to a log file, and the second one is a parser which reads the experiment log and calculates the statistics of the received packets in an offline manner. 

\subsection{Interference: Bluetooth}

Among the two types of interference considered in this paper, Bluetooth system is more prevalent as many of modern cars have Bluetooth-capable on-board stereo systems, and Bluetooth headsets are typically used by drivers. In order to emulate the realistic use case, a Bluetooth headset is placed inside the cabin, and a smartphone is placed above the center console to emulate the Bluetooth transceiver of the on-board stereo system. To generate constant interference, the smartphone continuously streams music to the Bluetooth headset.

The Bluetooth headset used here is a Sony DR-BT50 headset, and the smartphone used in the experiments is an Apple iPhone 4. Both of the devices support Bluetooth v2.0 standard or above, and hence they support adaptive frequency hopping scheme. In other words, the devices can dynamically change the frequency hopping sequence during the communication to avoid interference.

\subsection{Interference: WiFi}
Most of the in-vehicle WiFi networks in today's market are used to provide Internet access to the passengers. The system normally consists of a WiFi router and a cellular network gateway. To emulate an in-vehicle WiFi network, a Linksys WRT54G2 WiFi router is installed under the rear deck of the vehicle. A laptop is connected to the WiFi router through Ethernet to emulate a remote server. There is another laptop placed inside the cabin to be a WiFi client, which is associated with the Basic Service Set (BSS) of the WiFi router. To generate realistic WiFi traffic, a File Transfer Protocol (FTP) session is set up between the server and the client. FTP is a protocol that is widely used on the Internet to provide file transfer and resource sharing. Actually, the most common use case of the in-vehicle WiFi is web browsing, but a FTP session is used here to represent a worst case scenario to the IVWSNs (i.e., continuous WiFi interference).

Filezilla Server v0.9.41 runs on the laptop as the FTP server and Filezilla Client v3.5.3 is the FTP client on the WiFi client. Furthermore, to make the WiFi traffic more realistic, the downlink speed at the FTP server is limited at 800 kbps, which is a possible download speed of 3G networks\footnote{Most of the in-vehicle WiFi system uses 3G networks to connect to the Internet.}. Throughout the experiments, the FTP client will keep downloading large files from the FTP server to generate WiFi interference to the IVWSNs. Note that the WiFi devices use WiFi channel 6 (2437 MHz) in order to create interference to the ZigBee network, which uses ZigBee channel 17 (2435 MHz) in the experiments.

\section{Experimental Setup}
\begin{figure}[tbp]
\centering
\subfigure[Sensor location configuration 1: engine to engine]{
\includegraphics[scale = 0.45]{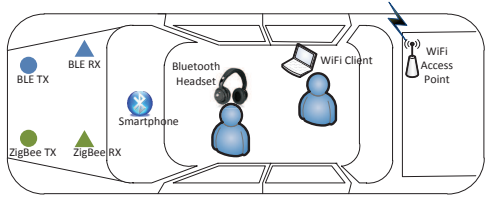}
\label{nodeconfig:subfig1}
}
\subfigure[Sensor location configuration 2: engine to cabin]{
\includegraphics[scale = 0.45]{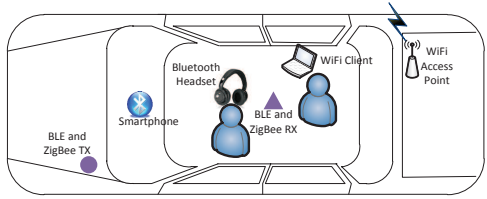}
\label{nodeconfig:subfig2}
}
\subfigure[Sensor location configuration 3: cabin to engine]{
\includegraphics[scale = 0.45]{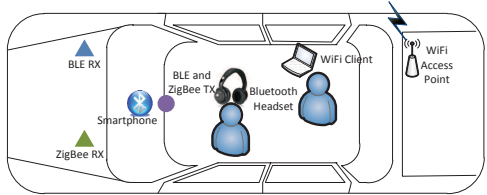}
\label{nodeconfig:subfig3}
}
\subfigure[Sensor location configuration 4: cabin to cabin]{
\includegraphics[scale = 0.45]{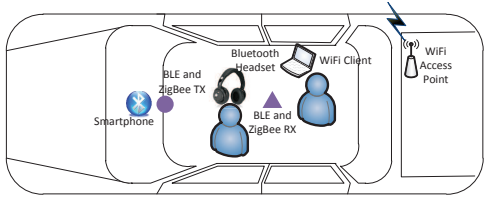}
\label{nodeconfig:subfig4}
}
\caption[]{The four sensor location configurations}
\label{NodeConf}
\end{figure}

The experiments were performed in a 2008 Chevrolet Impala, which is a common full-size sedan in the United States. To evaluate the performance of the two IVWSNs under interference, we designed comprehensive experiments that involve 48 different scenarios. Each scenario differs depending on the following four parameters (i.e., $2\times 4\times 3\times 2 = 48$):

\begin{enumerate}
\item Type of IVWSNs: as mentioned earlier, a total of two types of IVWSNs (i.e., BLE-based and ZigBee-based) are considered.
\item Location of sensors: since effect of interference may vary according to the relative locations of sensors and the interference sources inside the car, a total of four different sensor location configurations are considered: 
\begin{enumerate}
\item Engine to engine: the transmitter and the receiver are both within the engine compartment.
\item Engine to cabin: the transmitter is in the engine compartment, and the receiver is in the passenger compartment.
\item Cabin to engine: the transmitter is in the passenger compartment, and the receiver is in the engine compartment.
\item Cabin to cabin: the transmitter and the receiver are both within the passenger compartment.
\end{enumerate}
The four sensor location configurations are illustrated in Figure~\ref{NodeConf}. In each sensor location configuration, the sensor nodes are installed at the positions which are close to the actual sensor locations in the vehicle.
\item Type of interference: in our experiments, the effect of WiFi and Bluetooth interference were accounted for. Note that another set of experiments without any interference was also performed for benchmarking purpose. 
\item Type of environment: since it has been shown that the IVWSNs may behave differently  when the car is parked and is driven, in order to comprehensively study the effect of interference, we investigate such effect on both parking and driving scenario. For the driving scenario, the experiment was performed in Schenley Park, which is close to the CMU main campus. There are one driver and one passenger sitting in the middle of the rear seat; both driver and passenger have normal body movement during the entire experiment. For location/speed logging and time reference purpose, a GPS was also used.
\end{enumerate}

\begin{figure}[tbp]
\centering
\includegraphics[scale = 0.36, trim = 15mm 2mm 10mm 5mm, clip]{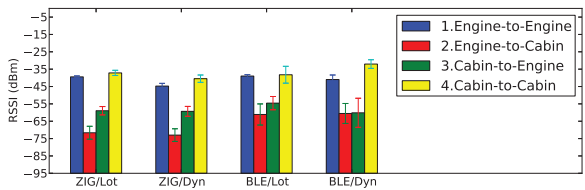}
\caption{The average RSSI of the two IVWSNs in each sensor location configuration}
\label{RSSIbar}
\end{figure}

For each of the 48 scenarios, the experiment was carried out for 5 minutes (i.e., at a packet rate of 4 Hz, a total of 1200 packets were transmitted and their statistics were collected). The Average Received Signal Strength Indicator (RSSI) along with its standard deviation (indicated by error bars) for each of the four sensor location configurations are presented in Figure~\ref{RSSIbar}. The x-axis of the figure specifies the types of IVWSNs and environment considered. For instance, ZIG/Lot indicates that the results were obtained from a parking lot scenario with ZigBee-based IVWSNs; BLE/Dyn indicates the results were from a dynamic driving scenario with BLE-based IVWSNs.

\begin{figure*}[tbp]
\centering
\subfigure[The WSN receivers]{
\includegraphics[scale = 0.23]{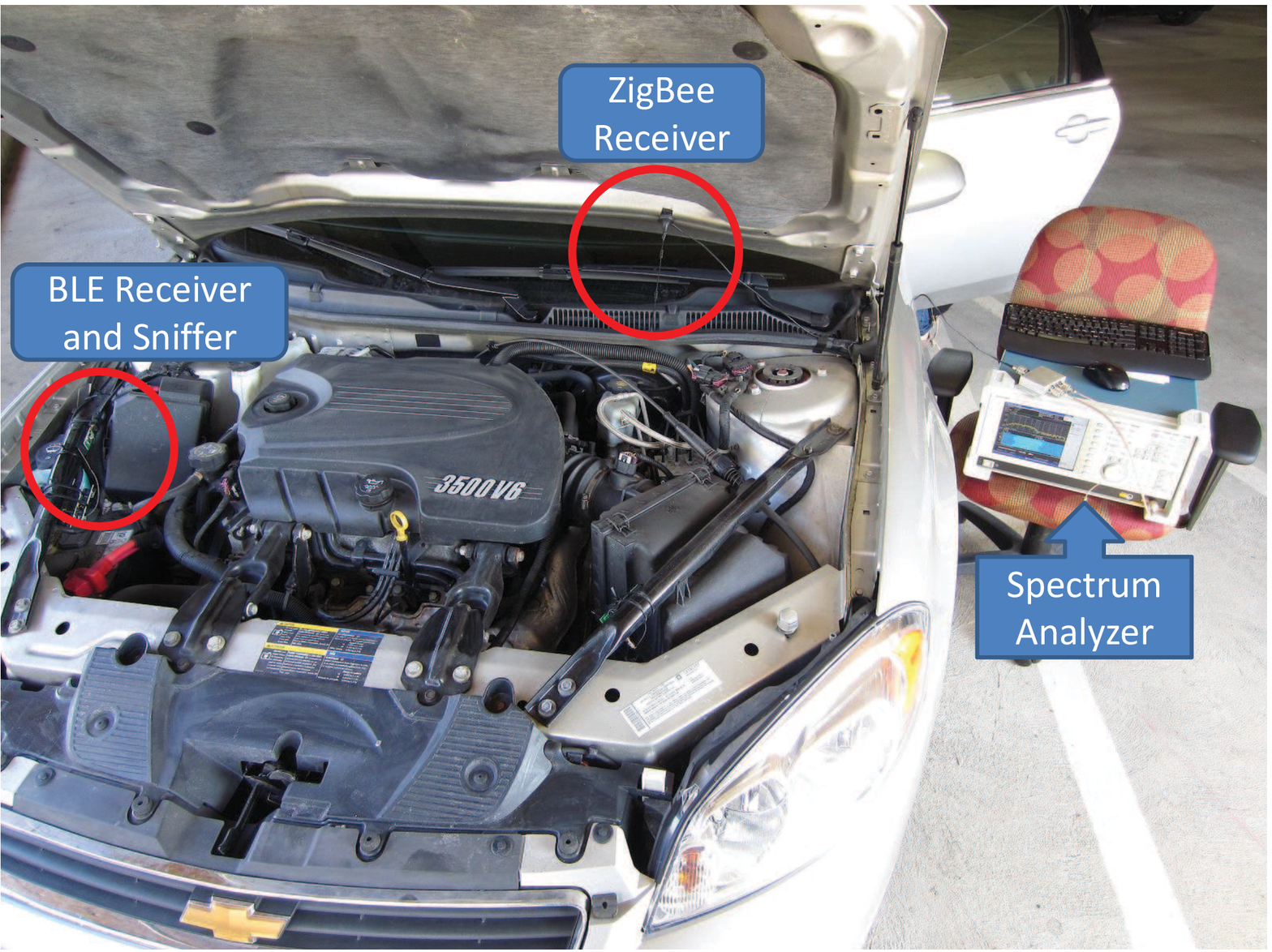}
\label{CP2_exp1}
}
\subfigure[The ZigBee transmitter]{
\includegraphics[scale = 0.25]{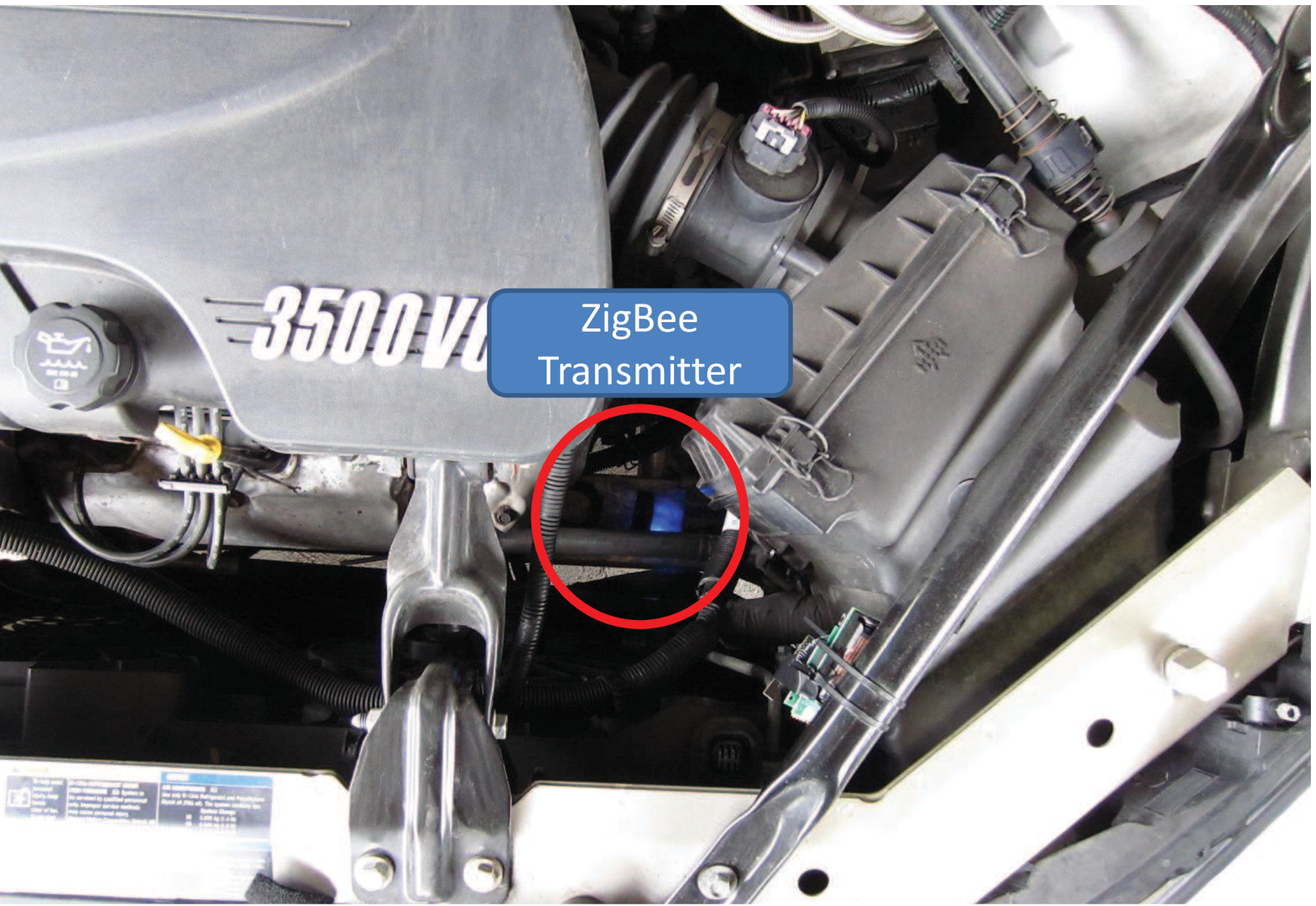}
\label{CP2_exp2}
}
\subfigure[The BLE transmitter]{
\includegraphics[scale = 0.25]{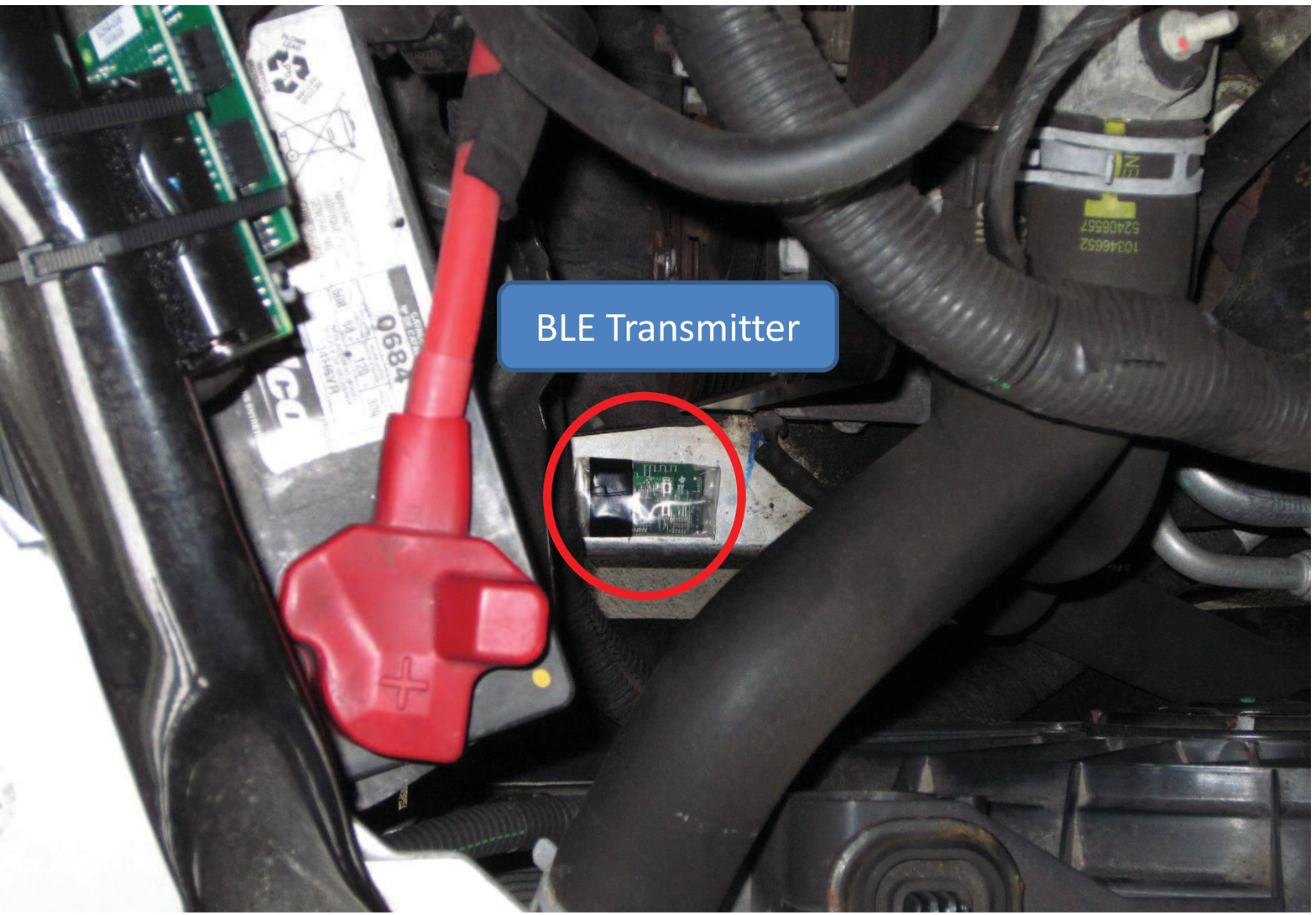}
\label{CP2_exp3}
}
\caption{The IVWSN transmitter/receiver locations of the sensor location configuration 1 (engine-to-engine)}
\label{CP2_NodeConf1}
\end{figure*}

In the experiments which interference is introduced, the average interference power at the WSN receivers is measured using a real-time spectrum analyzer in order to calculate the average Signal-to-Interference power Ratio (SIR). As an example, Figure~\ref{CP2_NodeConf1} shows the locations of the ZigBee and BLE transmitters and receivers in the sensor location configuration 1 (i.e., engine-to-engine), which corresponds to the setup shown in Figure~\ref{nodeconfig:subfig1}. %

The average interference power of WiFi signal is calculated by averaging the channel power of 10000 triggered signal samples, and the triggered threshold is set at -80 dBm. The Bluetooth interference power is calculated by averaging the channel power of 10 frequency hopping components. 
The average measured interference power  (i.e., of both WiFi and Bluetooth interference at the four receiving sensor locations shown in Figure~\ref{NodeConf}) are shown in Table~\ref{Interference_power}. Observe that ZigBee and BLE IVWSNs receive comparable amount of interference (caused by either WiFi or Bluetooth) when their receivers are in the same compartment. Moreover, much lower interference power is observed at the sensor located in the engine compartment, and this is mainly because the interference sources are located in either the passenger or trunk compartment.

\begin{table}[tbp]
  \centering
  \caption{The average interference power at each receiver position}
    \begin{tabular}{rcc}
    \toprule
          & Engine & Cabin \\
    \midrule
    WiFi interference power to ZigBee receiver & -72 dBm & -40 dBm \\
    Bluetooth interference power to ZigBee receiver & -55 dBm & -37 dBm \\
    WiFi interference power to BLE receiver & -71 dBm & -44 dBm \\
    Bluetooth interference power to BLE receiver & -58 dBm & -34 dBm \\
    \bottomrule
    \end{tabular}%
  \label{Interference_power}%
\end{table}%

\section{Experimental Results}

\begin{figure}[tbp]
\centering
\subfigure[Under Bluetooth interference]{
\includegraphics[scale = 0.36, trim = 15mm 1mm 10mm 5mm, clip]{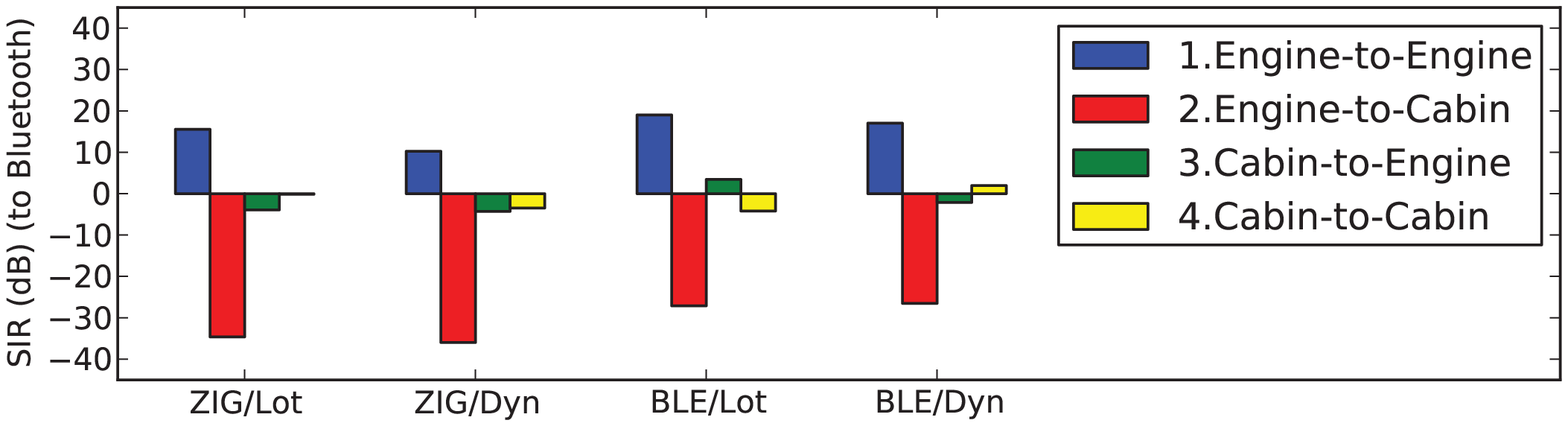}
\label{SIRbar:1}
}
\subfigure[Under WiFi interference]{
\includegraphics[scale = 0.36, trim = 15mm 1mm 10mm 5mm, clip]{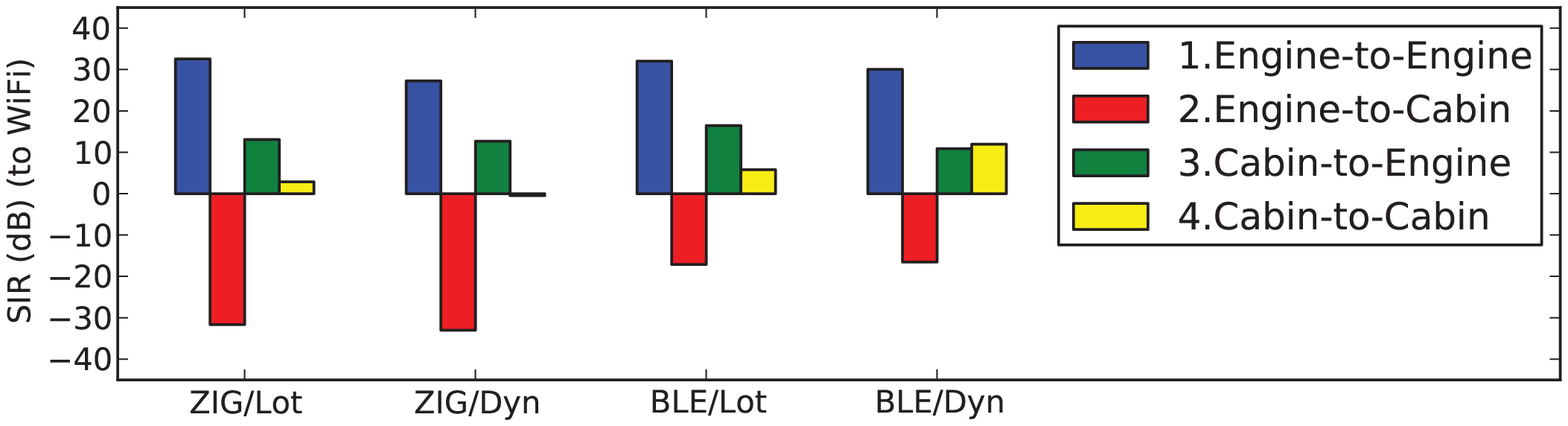}
\label{SIRbar:2}
}
\caption[]{The average SIR of each sensor location configuration}
\label{SIRbar}
\end{figure}

Based on the measured average received power at IVWSN sensors and average interference power of both WiFi and Bluetooth interference, one can compute the average Signal-to-Interference power Ratio (SIR) as follows:
\begin{align*}
\mbox{Average SIR} &= \frac{\mbox{Average RSSI of the received packets}}{\mbox{The average interference power (Table~\ref{Interference_power})}}
\end{align*}
The computed average SIR observed under two different interferences are shown in Figure~\ref{SIRbar}. Observe from the figure that among the four sensor location configurations, the engine-to-cabin configuration has the lowest SIR; and this is consistent with the observation made in the previous section. In addition, it is worth pointing out that the SIR values under WiFi interference are generally higher than those under Bluetooth interference. In order to analyze implications of the aforementioned observations on the performance of the IVWSNs, we use the goodput metric which measures the percentage at which the transmitted sensor packets are successfully received and decoded at the receiver. To be specific, the goodput metric can be defined as follows:
\begin{align*}
\mbox{Goodput }G  & \triangleq \frac{R-E}{T}\mbox{ ,}
\end{align*}
where T is the total number of packets transmitted by the transmitter, R is the number of packets received by the receiver, and E is the number of the received packets which fail the CRC check. 
Since we are interested in studying the performance of the ZigBee and BLE IVWSNs and the impact of the interference from WiFi and Bluetooth, we will compare the performance of the two IVWSNs under the following two assumptions.

\subsection{No interference case}

\begin{figure}[tbp]
\centering
\includegraphics[scale = 0.38]{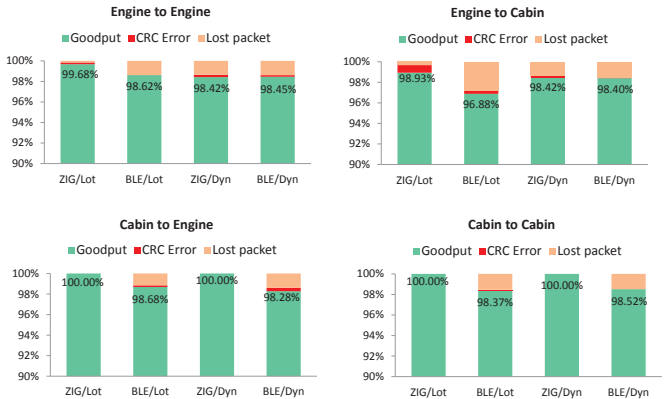}
\caption{The performance of BLE and ZigBee IVWSNs when no interference is introduced}
\label{result_1}
\end{figure}

In this subsection, a set of experiments when no interference exists was performed and the results of these experiments serve as a benchmark. Figure~\ref{result_1} shows the goodput performance (indicated by green bars) as well as the percentage of lost and erroneous packets (indicated by the orange and red bars, respectively). Four sets of figures are shown for each of the four different sensor location configurations; in each configuration, results of both types of IVWSNs and both parking (i.e., \textit{Lot}) and driving (i.e., \textit{Dyn}) scenarios are presented. Observe from the figure that when no interference exists, both ZigBee-based and BLE-based IVWSNs perform reasonably well with more than 96\% goodput. The ZigBee-based IVWSN performs slightly better than the BLE-based IVWSN and this is mainly due to the fact that the variance of the received signal power of BLE-based IVWSNs is larger then that of ZigBee-based system (see Figure~\ref{RSSIbar}). Higher fluctuation in terms of received power signal leads to slightly lower packet goodput (i.e, in most cases, the difference is less than 2\%).

\subsection{Bluetooth or WiFi interference}

\begin{figure}[tbp]
\centering
\includegraphics[scale = 0.47]{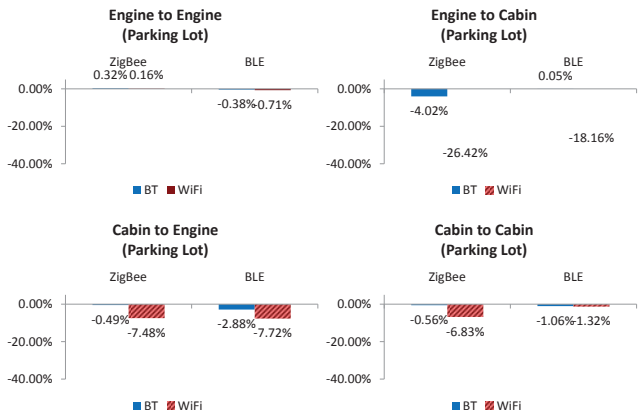}
\caption{Goodput degradation of BLE and ZigBee IVWSNs under interference (parking lot environment)}
\label{result_2}
\end{figure}

\begin{figure}[tbp]
\centering
\includegraphics[scale = 0.47]{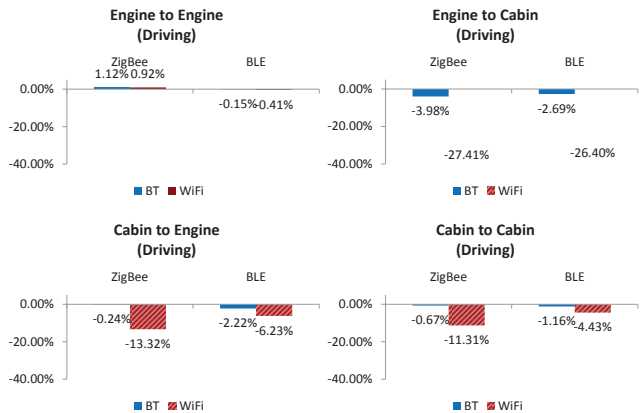}
\caption{Goodput degradation of BLE and ZigBee IVWSNs under interference (driving environment)}
\label{result_3}
\end{figure}

Using results of Figure~\ref{result_1} as a benchmark, Figure~\ref{result_2} and Figure \ref{result_3} present the percentage of \textit{goodput degradation} under Bluetooth (indicated by blue bars) or WiFi (indicated by red bars) interference in the parking and driving environments, respectively. Similar to Figure~\ref{result_1}, four sets of results are presented for each sensor location configuration. Observe that similar trends in terms of goodput degradation can be observed in both parking and driving environments with a slightly larger degradation in the driving environment. 

In addition, observe from the figure that no degradation in terms of goodput when the engine-to-engine configuration is considered. This is not surprising due to the fact that even under interference, the measured SIR at the receivers are still high (i.e., more than 10 dB for Bluetooth and 25 dB for WiFi). As a result, engine-to-engine sensor transmissions on both ZigBee-based and BLE-based IVWSNs are not affected by the interference. 

Besides the engine-to-engine configuration, because of much lower measured SIR values, noticeable goodput degradation can be observed in the other three sensor location configurations especially under the WiFi interference. In other words, up to 28\% goodput degradation is a result of WiFi interference while the Bluetooth interference leads to less than 5\% degradation. This observation may be counter-intuitive given that the SIR values under WiFi interference is larger than those under Bluetooth interference. For instance, consider the parking lot environment, ZigBee-based IVWSN, and the engine-to-cabin configuration, the WiFi's 26\% (as compared to Bluetooth's 4\%) goodput degradation is observed even though the measured SIR value under WiFi interference is 4 dB larger (i.e., -32 dB and -36 dB under WiFi and Bluetooth interference, respectively).

The above observation is a result of difference in underlying physical layer operations of WiFi and Bluetooth systems. In other words, WiFi uses the Direct Sequence Spread Spectrum (DSSS) with 22 MHz channel bandwidth whereas the Frequency Hopping Spread Spectrum (FHSS) with 1MHz channel bandwidth is used in Bluetooth devices. It therefore follows that the Bluetooth interference is \textit{more random} as compared to WiFi; the interference from Bluetooth only affects the ongoing transmission if it "hops" into the same frequency bands used by the IVWSNs.  It is worth pointing out that an adaptive FHSS is also supported where the Bluetooth device dynamically chooses the frequency band to further avoid causing interference. 

Another interesting observation is: when one compares the performance of two IVWSNs, we can observe that BLE-based IVWSN performs generally better than the ZigBee-based system and this phenomenon again arises due to the fact that the BLE sensor employs the FHSS technique whereas the ZigBee-based system does not. To conclude, based on our extensive experimental study, it is shown that the BLE-based IVWSNs considerably outperforms the ZigBee-based IVWSNs in terms of goodput when WiFi interference is introduced; and comparable performance is observed under Bluetooth interference.

\section{Conclusion}

In this paper, a detailed and comprehensive experimental study was performed to investigate the effect of Bluetooth and WiFi interference on the performance of the Intra-Vehicular Wireless Sensor Networks (IVWSNs). The experiments were setup to emulate realistic use cases of both Bluetooth and WiFi devices inside the vehicle. The results of a total of 48 different experimental scenarios suggest that both ZigBee-based and BLE-based IVWSNs (regardless of whether the car is parked or driven) perform reasonably well (in terms of goodput) when no interference or only Bluetooth interference are considered. The performance of both IVWSNs however significantly degrades when WiFi interference is introduced. Nevertheless, the BLE-based IVWSN is considerably more robust than the ZigBee-based network in most of the configurations. Thus, with the current technology, the BLE seems to be a better candidate for IVWSNs when robustness against interference is a main concern.

\ifCLASSOPTIONcaptionsoff
  \newpage
\fi

\linespread{0.85}
\bibliographystyle{IEEEtran}
\bibliography{int_arxiv}
\end{document}